\documentclass[aps,prl,twocolumn,groupedaddress,showpacs]{revtex4}
\usepackage{graphicx}
\usepackage{dcolumn}
\usepackage{bm}

\begin{document}

\title{Time in quantum gravity by weakening the Hamiltonian constraint}

\author{Hrvoje Nikoli\'c}
\affiliation{Theoretical Physics Division, Rudjer Bo\v{s}kovi\'{c} 
Institute,
P.O.B. 180, HR-10002 Zagreb, Croatia.}
\email{hrvoje@thphys.irb.hr}

\date{\today}

\begin{abstract}
We replace the usual Hamiltonian constraint of quantum gravity
$H|\psi\rangle=0$ by a weaker one $\langle\psi|H|\psi\rangle=0$.
This allows $|\psi\rangle$ to satisfy the time-dependent 
functional Schr\"odinger equation. In general, only the phase 
of the wave function appears to be time independent. 
The resulting quantum theory has the correct                 
classical limit and thus provides a viable theory of quantum 
gravity that solves the problem of time without introducing 
additional nongravitational degrees of freedom.
\end{abstract}


\maketitle

The classical Hamiltonian of the Einstein theory of gravity vanishes
when the metric satisfies the Einstein equations
\cite{mtw,kuc,padm,ish}. 
This is a consequence of the fact that the Einstein theory of gravity 
is a 
theory invariant with respect to arbitrary time reparametrizations.
It is widely believed that this classical Hamiltonian constraint
\begin{equation}\label{hamcl}
H=0
\end{equation}
should be quantized such that every physical state $\psi$ 
satisfies the 
Wheeler-DeWitt (WDW) \cite{dw} equation $\hat{H}\psi=0$, 
where $\hat{H}$ is the corresponding
quantum mechanical operator.  
On the other hand, the functional 
Schr\"odinger equation takes the form 
\begin{equation}\label{sch}
\hat{H}\psi=i\hbar\partial_t\psi. 
\end{equation}
Owing to the WDW equation, the right-hand side of (\ref{sch}) should 
vanish, which implies that $\psi$ does not depend on time. 
However, if $\psi$ does not depend on time, then it is not clear 
how it can be consistent with the fact that
classical quantities do depend on time. 
This is the problem of time 
in quantum gravity \cite{padm,ish}. 

The problem can be partially resolved when the interaction of gravity 
with matter is introduced. 
For example, solving the WDW equation
by using the WKB approximation, 
one finds that the matter wave functional satisfies a 
``many-fingered time" Tomonaga-Schwinger equation \cite{ban,kie}. 
However, this does 
not solve the problem of time beyond the WKB approximation. In particular,
this does not lead to any time dependence in the case of pure gravity 
without matter, which is in contradiction with the classical 
theory of gravity.

In this Letter we propose a simple general solution of the problem of 
time in quantum gravity. For simplicity, we study the case of pure gravity
without matter, but the results are easily generalized to any theory 
with a vanishing classical Hamiltonian, including the case of gravity 
interacting with matter. The basic idea is to allow $\psi$ to depend 
on time by satisfying (\ref{sch}), but, at the same time, 
to require that this should not contradict the Hamiltonian 
constraint (\ref{hamcl}) in the classical limit. As a byproduct, 
we also find that this approach partially resolves another important          
problem of quantum gravity -- the problem of operator ordering.
It appears that the requirement of unitary time evolution of $\psi$
leads to a restriction on the operator ordering of the Hamiltonian.

Let $\psi(g,t)=\langle g|\psi(t)\rangle$ 
be a time-dependent wave function, where $g$ represents 
the canonical degrees of freedom. (In the case of gravity, $g$ is the
3-metric.) Instead of the WDW equation $\hat{H}|\psi\rangle=0$, we require 
a weaker form of the quantum Hamiltonian constraint: 
\begin{equation}\label{hweak}
\langle\psi|\hat{H}|\psi\rangle=0.
\end{equation}
This corresponds to the requirement that the average value of $\hat{H}$ 
vanishes, which is consistent with the classical Hamiltonian 
constraint (\ref{hamcl}). 

Note that such a weaker form of the 
constraint is similar to the Gupta-Bleuler quantization of the 
classical Lorentz-condition constraint $\partial_{\mu}A^{\mu}=0$ 
of electrodynamics \cite{ryd,schweber}. In quantum electrodynamics, 
it is not consistent to require 
$\partial_{\mu}\hat{A}^{\mu}|\psi\rangle=0$. However, one can write 
$\hat{A}^{\mu}=\hat{A}^{\mu(+)}+\hat{A}^{\mu(-)}$, 
where $\hat{A}^{\mu(+)}$ ($\hat{A}^{\mu(-)}$) is
the positive (negative) frequency part. The Gupta-Bleuler method 
consists in requiring $\partial_{\mu}\hat{A}^{\mu(+)}|\psi\rangle=0$, 
which is equivalent to $\langle\psi|\partial_{\mu}\hat{A}^{\mu(-)}=0$.
This implies $\langle\psi|\partial_{\mu}\hat{A}^{\mu}|\psi\rangle=0$, 
which is analogous to (\ref{hweak}). 

The condition (\ref{hweak}) can be written as
\begin{equation}\label{hweak2}
\int {\cal D}g\, \psi^*\hat{H}\psi=0.
\end{equation}
By writing $\psi=Re^{iS/\hbar}$ (where $R$ ans $S$ are real), 
using (\ref{sch}), and requiring 
the conservation of probability
\begin{equation}\label{unit}
\frac{d}{dt} \int{\cal D}g\, \psi^*\psi =0,
\end{equation}
one finds that (\ref{hweak2}) is fulfilled if
\begin{equation}\label{s}
\partial_t S=0.
\end{equation}
In other words, the Hamiltonian constraint implies that the wave 
function takes the form
\begin{equation}\label{wf}
\psi(g,t)=R(g,t)e^{iS(g)/\hbar}.
\end{equation}
The phase $S$ is time independent, but the square of the absolute value 
$|\psi|^2=R^2$ may depend on time.

Let us investigate the idea above in more detail.
To simplify the analysis, we write the Hamiltonian of classical 
gravity in the condensed notation
\begin{equation}\label{hcond}
H=G_{AB}(g)\pi^{A}\pi^{B}+V(g),
\end{equation}
where $g\equiv \{ g_A \}$ and $G_{AB}=G_{BA}$. Explicitly,
\begin{equation}\label{not1}
G_{AB}\pi^{A}\pi^{B}\equiv \kappa\int d^3x\, G_{ijkl}\pi^{ij}\pi^{kl},
\end{equation}
\begin{equation}\label{not2}
V\equiv -\kappa^{-1}\int d^3x\,\sqrt{|g^{(3)}|}R^{(3)}.
\end{equation}
Here 
$\pi^{ij}$ $(i,j=1,2,3)$ are the canonical momenta corresponding 
to the canonical coordinates $g_{ij}$, 
$g^{(3)}$ is the determinant of $g_{ij}$, $R^{(3)}$ is the curvature 
constructed from the metric $g_{ij}$, $\kappa=8\pi G_N$, 
and 
\begin{equation}
G_{ijkl}=\frac{\sqrt{|g^{(3)}|}}{2} 
(g_{ik}g_{jl}+g_{jk}g_{il}-g_{ij}g_{kl})
\end{equation}
is the supermetric.

In the quantum case, the momenta $\pi^A$ become the derivative 
operators
\begin{equation}\label{piop}
\hat{\pi}^A=-i\hbar\frac{\delta}{\delta g_A}\equiv -i\hbar\partial^A.
\end{equation}
Since $G_{AB}$ depends on $g$, different orderings of the operators 
in the first term of (\ref{hcond}) are not equivalent.
Therefore, we study a quantum variant of (\ref{hcond}) of the form
\begin{equation}\label{hcond2}
\hat{H}=\frac{1}{a+b}[aG_{AB}\hat{\pi}^{A}\hat{\pi}^{B}
+b\hat{\pi}^{A}G_{AB}\hat{\pi}^{B}] +V,
\end{equation} 
where $a$ and $b$ are arbitrary constants. Eq.~(\ref{hcond2}) 
certainly does not correspond to the most general ordering of $\hat{H}$,
but represents a large class of possible orderings. This class 
of orderings will be sufficient to demonstrate the fact that 
the unitarity condition (\ref{unit}) imposes restrictions on 
the otherwise arbitrary choice of ordering. Using (\ref{piop}), 
(\ref{hcond2}) becomes
\begin{equation}
\hat{H}=-\hbar^2[G_{AB}\partial^A\partial^B 
+cG_{AB}^{\;\;\;\;\; ,A}\partial^B ] +V,
\end{equation}
where $G_{AB}^{\;\;\;\;\; ,A}\equiv\partial^A G_{AB}$ and
$c=b/(a+b)$. Using (\ref{wf}), one finds that
the complex equation (\ref{sch}) 
is equivalent to a set of two real equations
\begin{equation}\label{qhj}
G_{AB}S^{,A}S^{,B}+V+Q=0,
\end{equation}
\begin{equation}\label{cons}
\partial_t R^2+\partial^A(R^2 2G_{AB}S^{,B})
+(c-1)2R^2 G_{AB}^{\;\;\;\;\; ,A}S^{,B} =0,
\end{equation} 
where
\begin{equation}\label{Q}
Q=-\hbar^2\left( G_{AB}\frac{R^{,AB}}{R} +cG_{AB}^{\;\;\;\;\;,B}
\frac{R^{,A}}{R} \right) .
\end{equation}
Now we see that (\ref{unit}) is satisfied if the last term 
in (\ref{cons}) vanishes, because then the left-hand side of 
(\ref{unit}) reduces to an integral over a total derivative, which 
vanishes. This implies that we must take $c=1$ which is equivalent 
to $a=0$, so (\ref{hcond2}) reduces to
\begin{equation}\label{hcond3}
\hat{H}=\hat{\pi}^{A}G_{AB}\hat{\pi}^{B} +V.
\end{equation}
This fixes a unique ordering among all the orderings that belong 
to the class of orderings defined by (\ref{hcond2}). Note that this
ordering provides that not only norms 
$\langle\psi|\psi\rangle$, but also all scalar products
$\langle\psi_1|\psi_2\rangle$, are time independent. This is because 
(\ref{sch}) and (\ref{hcond3}) imply 
\begin{equation}
\frac{d}{dt} \int{\cal D}g\,\psi_1^*\psi_2 =
\hbar \int{\cal D}g\, \partial^A[G_{AB}(\psi_1^*i
\!\stackrel{\leftrightarrow\;}{\partial^{B}}\!
\psi_2)],
\end{equation}
which vanishes because the integral over a total derivative vanishes. 

Let us now consider the classical limit $\hbar\rightarrow 0$.
Eqs.~(\ref{qhj}) and (\ref{cons}) with $c=1$ become
\begin{equation}\label{qhj2}
G_{AB}S^{,A}S^{,B}+V=0,
\end{equation}
\begin{equation}\label{cons2}
\partial_t R^2+\partial^A(R^2 2G_{AB}S^{,B})=0.
\end{equation}
Eq.~(\ref{qhj2}) is the classical Hamilton-Jacobi equation
equivalent to the classical Hamiltonian constraint (\ref{hamcl}),
provided that we make the classical identification
$\pi^A=S^{,A}$. The classical time derivative of the metric
$\dot{g}_A\equiv \partial_t g_A$ 
is given by the classical Hamilton equation
\begin{equation}\label{heq}
\dot{g}_A=\frac{\partial H}{\partial\pi^A}=2G_{AB}\pi^B,
\end{equation}
so Eq.~(\ref{cons2}) can be written as
\begin{equation}\label{cons3}
\partial_t \rho +\partial^A(\rho \dot{g}_A)=0,
\end{equation}
where $\rho=R^2$. Eq.~(\ref{cons3}) is naturally interpreted 
as a local (in the configuration space)
probability conservation in a classical 
statistical ensemble of metric configurations described by 
the probability density $\rho(g,t)$. This conservation can be written 
in terms of a total derivative as
\begin{equation}\label{constot}
\frac{d\rho}{dt}=0.
\end{equation}
Note that the analog of (\ref{cons3}) 
for the WDW equation corresponds to $\rho$ that does not 
explicitly depend on time, i.e., satisfies $\partial_t\rho=0$.
On the other hand, there is no reason for a classical probability density 
to satisfy this constraint. The initial classical probability density 
$\rho(g,t_0)$ is arbitrary, while its further time evolution is 
determined by classical dynamics, which, in general, leads to an 
explicitly time-dependent solution $\rho$ of Eq.~(\ref{cons3}).
In this sense, the conventional stronger 
form of the Hamiltonian constraint does not have the correct
classical limit, while our weaker form of the 
Hamiltonian constraint does. 

Note that the interpretation of $\rho$ 
as a classical probability density 
in the limit $\hbar\rightarrow 0$
is particularly natural from the point of view of the Bohmian 
interpretation of quantum gravity, which is usually formulated within 
the conventional WDW approach  
(see, e.g., \cite{shoj} and references therein). However, 
in order to interpret $\rho$ as a classical probability density
in the limit $\hbar\rightarrow 0$, 
it is not necessary to adopt the Bohmian interpretation of the 
full quantum theory.
 
For those who are skeptical about such a probabilistic interpretation
of $\rho$ in the classical limit, there are also other, 
more conventional, ways 
of looking at the classical limit. One possiblity is to construct 
a solution $\psi(g,t)$ peaked around a classical solution 
$g_A^{{\rm class}}(t)$, so that
\begin{equation}
R^2(g,t)\approx\delta(g-g^{{\rm class}}(t)),
\end{equation}
where $\delta(g)$ is a short symbol for $\prod_A \delta(g_A)$.
Then the average value of $g_A$ is
\begin{equation}
\langle g_A\rangle =\int{\cal D}g\, \psi^*(g,t)g_A\psi(g,t)
\approx g_A^{{\rm class}}(t).
\end{equation}
We see that it is necessary that $R$ depends on time in order 
to obtain the correct classical
limit with a time-dependent metric.
Another approach to the classical limit is through the Ehrenfest 
theorem. The Schr\"odinger equation (\ref{sch}) implies
\begin{eqnarray}\label{ehr}
& \partial_t \langle\psi| g_A|\psi\rangle=
\langle\psi| i\hbar^{-1}[\hat{H},g_A]|\psi\rangle, & \nonumber \\
& \partial_t \langle\psi| \hat{\pi}^A|\psi\rangle=
\langle\psi| i\hbar^{-1}[\hat{H},\hat{\pi}^A]|\psi\rangle. &
\end{eqnarray}
Using the fact that
\begin{eqnarray}
& i\hbar^{-1}[\hat{H},g_A]=\{ \hat{H},g_A\}_{{\rm PB}}+{\cal O}(\hbar), 
 & \nonumber \\ 
& i\hbar^{-1}[\hat{H},\hat{\pi}^A]=\{ \hat{H},\hat{\pi}^A\}_{{\rm PB}}
+{\cal O}(\hbar), &
\end{eqnarray}
we see that the classical limit $\hbar\rightarrow 0$ applied 
to (\ref{ehr}) leads to the classical equations of motion 
for the averaged values, provided that $|\psi\rangle$ satisfies 
a time-dependent Schr\"odinger equation (\ref{sch}).

So far, we have not discussed a very important issue -- the issue 
of reparametrization invariance of our quantum theory. For that 
purpose, note first that the Hamiltonian $H=\int d^3x{\cal H}$, 
known also under the name superhamiltonian,
should be distinguished from the canonical Hamiltonian
\begin{equation}
H^{{\rm can}}=\int d^3x [N{\cal H}+N_i{\cal H}^i],
\end{equation}
where $N=1/\sqrt{|g^{00}|}$ is the lapse function, $N_i=g_{0i}$ 
is the shift function, and ${\cal H}^i=-2\pi^{ij}_{\;\;\;|j}$ 
is the supermomentum density. 
Strictly speaking, one should 
replace (\ref{sch}) with 
\begin{equation}\label{fullsch}
\hat{H}^{{\rm can}}\psi=i\partial_t \psi .
\end{equation}
In this sense, our previous results 
correspond to the choice of gauge $N=1$, $N_i=0$, because 
only then $H=H^{{\rm can}}$. However, it is easy to modify
all the equations such that they are correct for all choices
of $N$ and $N_i$. 
Integrating by parts, one finds
\begin{equation}
\int d^3x\, N_i{\cal H}^i=
\int d^3x\, 2[N_{i|j}-\Gamma^k_{kj}N_i]\pi^{ij}
\equiv F_A\pi^A.
\end{equation}
Redefining $G_{AB}$ in (\ref{not1}) and $V$ in (\ref{not2}) such that 
the factor $N$ is included, the generalization of (\ref{hcond3}) is 
\begin{equation}\label{hcond3.1}
\hat{H}^{{\rm can}}=\hat{\pi}^{A}G_{AB}\hat{\pi}^{B} 
+\frac{1}{2}(F_A\hat{\pi}^A+\hat{\pi}^AF_A) +V.
\end{equation}
With this ordering, the generalizations of (\ref{qhj}) and (\ref{cons})
are
\begin{equation}\label{qhj.1}
G_{AB}S^{,A}S^{,B}+F_AS^{,A}+V+Q=0,   
\end{equation}
\begin{equation}\label{cons.1}
\partial_t R^2+\partial^A[R^2 (2G_{AB}S^{,B}+F_A)]=0.
\end{equation}
In the classical limit, (\ref{qhj.1}) becomes the
Hamilton-Jacobi equation for the classical constraint 
$H^{{\rm can}}=0$, while (\ref{cons.1}), together with 
$\pi^A=S^{,A}$ and the 
generalization of (\ref{heq}) 
\begin{equation}\label{heq.1}
\dot{g}_A=\frac{\partial H^{{\rm can}}}{\partial\pi^A}=2G_{AB}\pi^B+F_A,
\end{equation}
becomes the conservation equation (\ref{cons3}) equivalent to 
(\ref{constot}).
In addition, we also quantize the classical supermomentum
constraint ${\cal H}^i=0$ in the usual way, by postulating 
\begin{equation}\label{supmom}
\hat{{\cal H}}^i\psi=0.
\end{equation}
This constraint provides the reparametrization invariance of the quantum 
theory with respect to the coordinate transformations of the form 
$x^i\rightarrow x'^i=f^i(x^1,x^2,x^3)$. The constraint (\ref{supmom}) 
implies that (\ref{fullsch}) is equivalent to (\ref{sch}) with a redefined 
Hamiltonian
\begin{equation}
\hat{H}=\int d^3x\, N\hat{{\cal H}}.
\end{equation}
The resulting Schr\"odinger equation (\ref{sch}) is covariant with 
respect to time reparametrizations of the form $t\rightarrow t'(t)$.
The variables $N$ and $N_i$ are not quantized because they are not 
dynamical, i.e., because they do not possess the corresponding canonical 
momenta. 

We still do not have the full reparametrization invariance with 
respect to general coordinate transformations of the form
$x^{\mu}\rightarrow x'^{\mu}=f^{\mu}(x^0,x^1,x^2,x^3)$, 
$\mu=0,1,2,3$. We only have
the reparametrization invariance with respect to coordinate 
transformations that do not mix space coordinates with time coordinates,
i.e., with respect to transformations that preserve the foliation of 
spacetime into spacelike hypersurfaces $\Sigma$.
It is, of course, 
impossible to obtain a complete symmetry between space and time 
coordinates in a canonical approach. Even the classical canonical 
theory requires the global spacetime topology of the form 
$\Sigma\times{\bf R}$, which introduces asymmetry 
between space and time.
However, the WDW approach treats the space and time 
constraints in a more symmetrical way than we do, by postulating 
${\cal H}\psi=0$, ${\cal H}^i\psi=0$. In this sense, the WDW 
approach may seem more appealing than our approach. On the other hand, 
as we have seen, the WDW 
approach does not have the correct classical limit, while our approach 
does. If a classical theory is verified experimentally, then
it is more important for a quantum theory to have the correct 
classical limit  
than to possess all the symmetries that the classical 
theory possesses. Indeed, the existence of various quantum anomalies 
teaches us that the quantum 
theory does {\em not} need to possess 
all the symmetries that the classical theory does.

Note also that the existence of a preferred foliation of spacetime
is not necessarily a problem if one takes the viewpoint according
to which the
Einstein theory of gravity may correspond to an effective 
theory for an underlying more 
fundamental theory that possesses a preferred 
time even at the classical level. This speculation is  
supported by the fact that 
the low-energy phenomena of various nonrelativistic systems, such as 
certain kinds of fluids \cite{vis,vol},  
may be described by generally covariant equations.

Finally, let us note one new way of looking at the problem of 
general covariance in quantum gravity. The WDW equations for $\psi$
contain neither space nor time derivatives and therefore they are 
covariant. Our Schr\"odinger equation contains a time derivative 
but does not contain space derivatives, so it is not 
covariant. However, in our approach, $S$ does not depend on time, 
so (\ref{qhj}) and (\ref{qhj.1}) take the same covariant form 
as the analogous equations in the WDW approach. (If $S$ were time 
dependent, then the right-hand sides of (\ref{qhj}) and (\ref{qhj.1}) 
would be equal to $-\partial_t S$.) The nonvanishing partial 
time derivatives appear only in (\ref{cons}) and (\ref{cons.1}), 
which makes these equations noncovariant. However, there is 
a way of writing the noncovariant equation (\ref{cons.1}) in a covariant 
form. All we have to do is to assume that the classical equations 
(\ref{heq.1}) and $\pi^A=S^{,A}$ are, in some sense, also valid 
in the quantum case. With this assumption, (\ref{cons.1}) 
can be written as (\ref{cons3}) which is equivalent to (\ref{constot}).
Since $\rho$ does not depend on $x^i$ (recall that $\psi$ is a functional, 
not a function, of $g_{ij}(x^1,x^2,x^3)$),
(\ref{constot}) can be written in the covariant form as
\begin{equation}
\frac{d\rho}{d x^{\mu}}=0.
\end{equation}
Fortunately, there is a way of making sense of equations
(\ref{heq.1}) and $\pi^A=S^{,A}$ in the quantum domain; they
correspond to the already mentioned Bohmian interpretation of 
quantum gravity. In this interpretation, $g_A(t)$ is a deterministic 
hidden variable.
Note that, owing to the nonclassical term $Q$ in (\ref{qhj.1}), 
the equation of motion (\ref{heq.1}) is not covariant; 
the dynamics of $g_{\mu\nu}(x^0,x^1,x^2,x^3)$  
is described by a modification of the classical 
Einstein equation covariant only with respect to  
transformations that preserve the foliation of spacetime
\cite{shoj}. In a way, the noncovariance is translated 
from the equations that determine the wave function $\psi$ to the 
equations that determine unobservable hidden variables. In other 
words, we have a remarkable and surprising result that the Bohmian 
interpretation should be invoked in order to write 
the equations that determine $\psi$ in a covariant form. It remains 
to be seen in future research if this result has deeper consequences
on the role of the Bohmian interpretation in quantum gravity.  

To summarize, in this Letter we have shown that time 
in pure quantum gravity 
without additional nongravitational degrees of freedom can be introduced 
by replacing the usual Hamiltonian constraint $\hat{H}|\psi\rangle=0$
by a weaker one $\langle\psi|\hat{H}|\psi\rangle=0$.
The results are easily generalized to any theory
with a vanishing classical Hamiltonian, including the case of gravity
interacting with matter.
Contrary to the conventional approach, based on the WDW equation, 
which cannot explain why metric can depend on time in the classical 
limit, our approach does explain that.
Unfortunately, with our approach, we do not completely solve the
problem of the full reparametrization invariance in quantum gravity,
but at least we provide some hints for a solution of this problem. 
In particular, it appears that the Bohmian interpretation
might have a previously unnoted role in establishing the 
general covariance of quantum gravity.

This work was supported by the Ministry of Science and Technology of the
Republic of Croatia under Contract No.~0098002.

\end{document}